\documentclass[12pt]{book}

\usepackage[dvips]{graphicx,color}
\usepackage{wrapfig}
\usepackage{makeidx,tsukuba}

\makeauthorindex
\makeindex

\begin{document}

\BookTitle{\itshape The 28th International Cosmic Ray Conference}
\CopyRight{\copyright 2003 by Universal Academy Press, Inc.}
\pagenumbering{arabic}

\chapter{
Calculation of Muon Fluxes at the Small Atmospheric Depths}

\author{%
%
%
K. Abe,$^1$ M. Honda,$^2$ K. Kasahara,$^3$ T. Kajita,$^2$ 
S. Midorikawa,$^4$ T. Sanuki,$^5$\\
{\it 
(1) Kobe University, Kobe, Hyogo 657-8501, Japan\\ 
(2) Institute for Cosmic Ray Research, The University of Tokyo,
Kashiwa, Chiba 277-8582, Japan\\
(3) Shibaura Institute of Technology, Ohmiya, Saitama
 330-8570, Japan\\
(4) Faculty of Engineering, Aomori University, Aomori, Aomori
030-0943, Japan\\
(5) The University of Tokyo, Bunkyo, Tokyo 113-0033, Japan
} 
}

\section*{Abstract}
Precise knowledge of the hadronic interaction between 
primary cosmic rays and atmospheric nuclei is 
very important and fundamental to 
study atmospheric neutrinos and their oscillations. 
We studied hadronic interaction models using the data of primary 
and secondary cosmic rays observed 
by BESS experiments.
By comparing the observed spectra with the ones calculated  
by several interaction models, 
we found DPMJET-III is most favored among them.

\section{Introduction}

The discovery of neutrino oscillations and its finite mass 
using the atmospheric 
neutrino is a mile stone in the history of particle physics.
The next step would be the accurate determination of the oscillation parameters.
However, the capability of neutrino experiments using the atmospheric neutrinos
is limited by the accuracy of the predicted neutrino fluxes.
The main uncertainty in the calculation of the atmospheric 
neutrino flux stems from the uncertainties of primary cosmic-ray 
flux and hadronic interactions.
We note that the uncertainty of the cosmic-ray proton fluxes is remarkably 
reduced by the recent cosmic ray observations up to 100~GeV.
For the hadronic interactions, however, there are scarcely any  
recent experiments available for our purpose.

In this paper, we study interaction models using 
the data of primary 
and secondary cosmic rays observed simultaneously 
by the BESS experiment at balloon altitudes. 
Muons at balloon altitudes are considered to carry direct 
informations of the hadronic interaction of primary cosmic rays with 
air nuclei. However, it is usually difficult to acquire sufficient 
statistics during balloon flights, 
due to small muon fluxes at balloon altitudes.
The BESS-2001 flight is unique and interesting in this sense. 
During the 
flight, the balloon kept relatively lower altitudes, corresponding
to the air depths of 4.5 - 28 ${\rm g/cm^2}$ for a long time,
and registered a sufficient number of muons.
This paper is a revise of the previous work~[2]
after the calibration of the atmospheric depth measurement.

\section{BESS Experiments}
The BESS ($\underline{\rm B}$alloon-borne $\underline{\rm E}$xperiment
with a $\underline{\rm S}$uperconducting 
$\underline{\rm S}$pectrometer)
detector~[3,4,9,14,15] 
is a high-resolution spectrometer with a large acceptance
to perform precise measurement of absolute fluxes
of various primary cosmic rays,  
as well as highly sensitive searches
for rare cosmic-ray components. 
In the previous measurements, BESS obtained 
precise atmospheric muon 
spectra at sea level~[7] and mountain 
altitudes~[13]
as well as precise primary proton and helium 
spectra~[12].

\begin{wrapfigure}[8]{r}[0pt]{6.5cm}
\vspace{-1.2cm}
  \begin{center}
    \includegraphics[width=6.5cm]{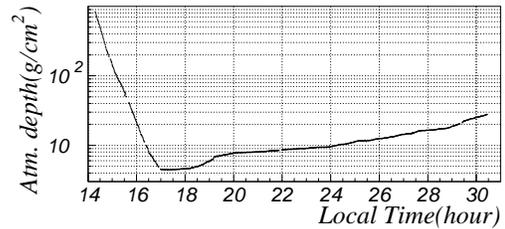}
  \end{center}
  \vspace{-0.5cm}
  \caption{Atmospheric depths during the BESS-2001 flight. }
  \label{fig:monitor}
\end{wrapfigure}

The BESS-2001 balloon flight was carried out at Ft. Sumner, 
New Mexico, 
USA (34$^{\circ}$49$'$N, 104$^{\circ}$22$'$W) on 24th September 
2001. 
Throughout the flight, the vertical geomagnetic cut-off rigidity 
was about 4.2~GV. 
Figure 1 shows a balloon flight 
profile during the experiment. 
The balloon reached at a normal floating altitude of 36~km at an 
atmospheric depth of 4.5~g/cm$^2$, then 
gradually lost its altitude. 
During the descending period, cosmic-ray data were collected at 
atmospheric depths between 4.5 g/cm$^2$ and 28 g/cm$^2$.

\begin{figure}[b!]
\begin{minipage}{7cm}
\vspace{-0.8cm}
  \begin{center}
    \includegraphics[width=6.55cm]{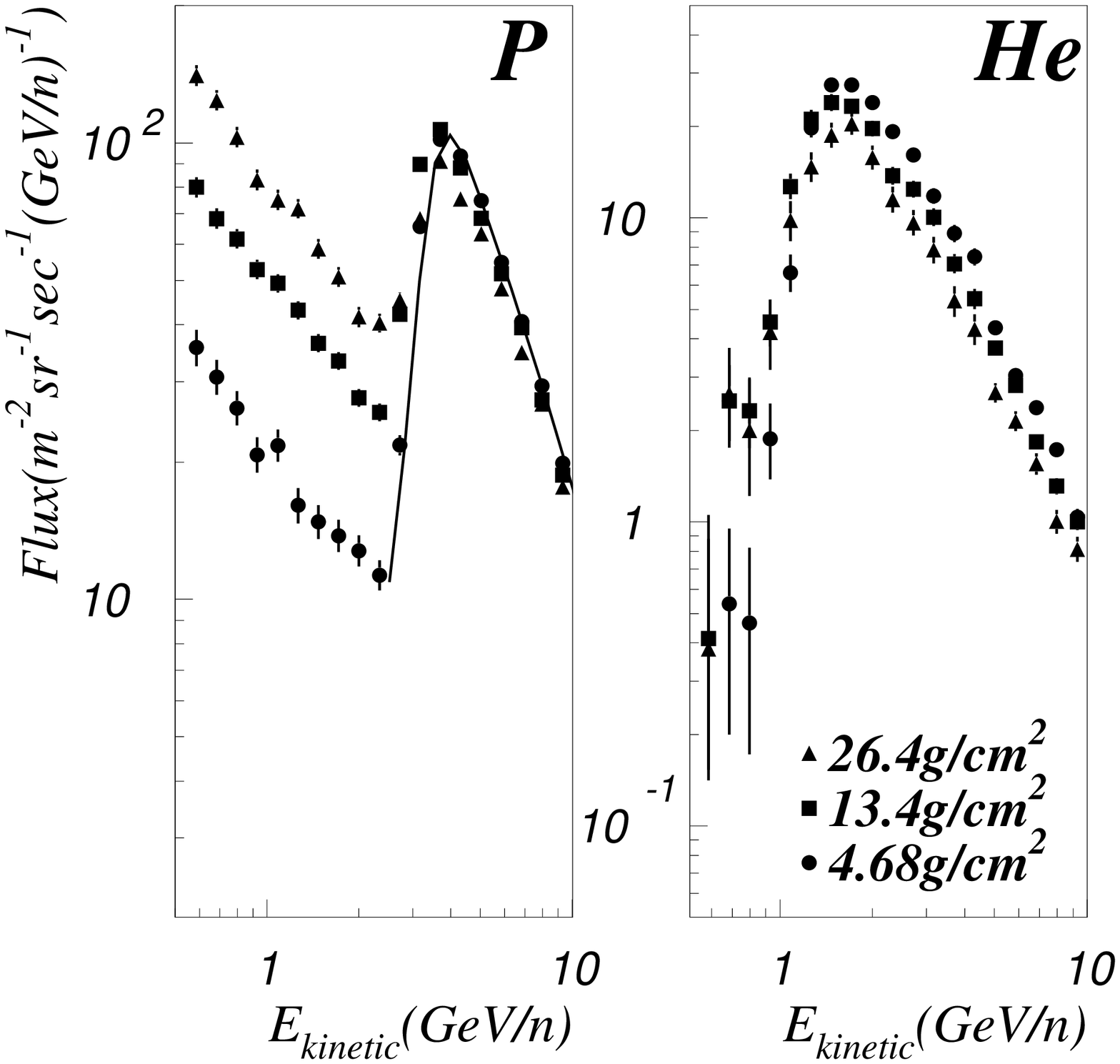}
  \end{center}
  \vspace{-0.5cm}
  \caption{The observed proton and helium spectra. 
    Solid line shows the calculated 
    proton fluxes at 4.68~g/cm$^2$.}
  \label{fig:pheflux}
\end{minipage}
\hspace{0.5cm}
\begin{minipage}{7cm}
\vspace{-1.cm}
  \begin{center}
    \includegraphics[width=6.55cm]{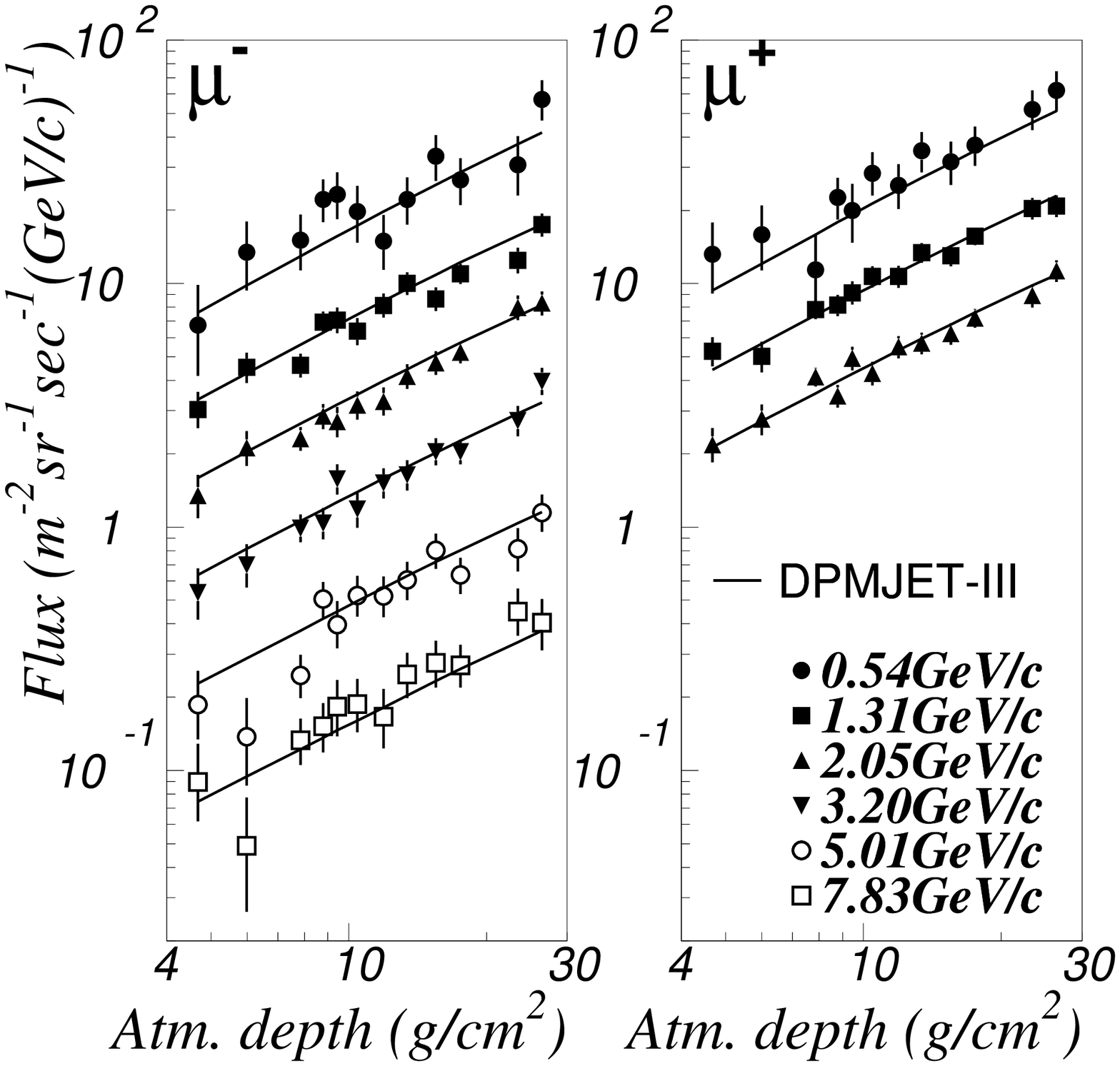}
  \end{center}
  \vspace{-0.5cm}
  \caption{The observed negative and positive 
    muon fluxes as a function of atm. depth.}
  \label{fig:diffintm}
\end{minipage}
\end{figure}

The proton and helium fluxes in the energy range of 0.5--10~GeV/n
and muon flux in 0.5~GeV/$c$--10~GeV/$c$ 
were obtained~[1]. 
The overall errors including both statistic and systematic 
errors are less than 8~\%, 10\% and 20~\% 
for protons, helium nuclei and muons, respectively. 
The obtained proton and helium spectra are 
shown in Fig.\ 2. 
Figure 3 shows the observed 
muon flux together with the predicted ones~[11] 
as a function of the residual atmospheric depth. 
For understanding the interactions 
and tuning the models in the calculation, 
these data are essentially important to 
be compared with calculations. 

\begin{wrapfigure}[30]{r}[0pt]{9.7cm}
\vspace{-1.3cm}
  \begin{center}
    \includegraphics[width=9.7cm]{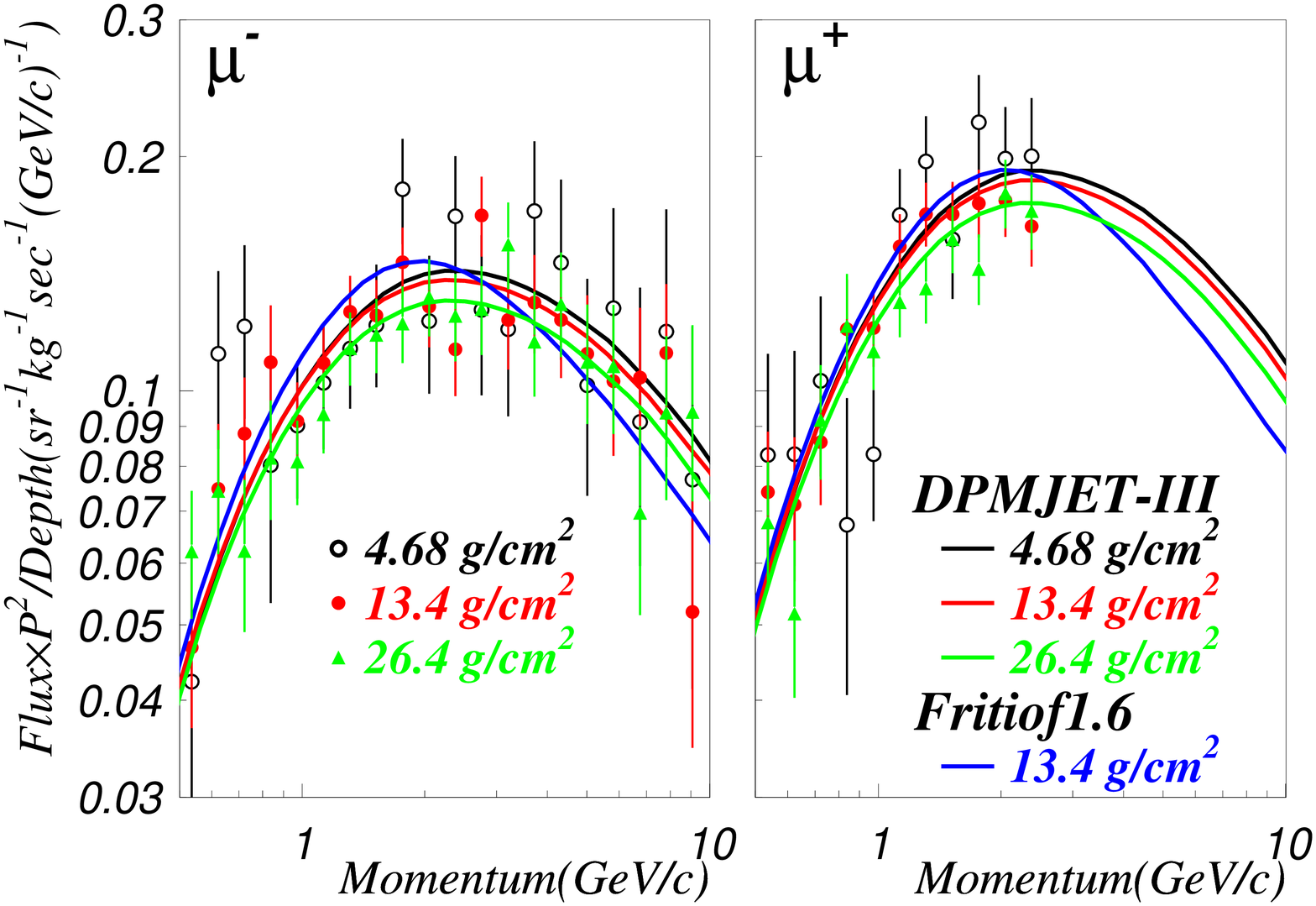}
  \end{center}
  \vspace{-0.6cm}
  \caption{The observed and calculated 
    (muon flux)/(atm. depth in kg/m$^2$).}
  \label{fig:compar}
\vspace{-1.0cm}
  \begin{center}
    \includegraphics[width=9.7cm]{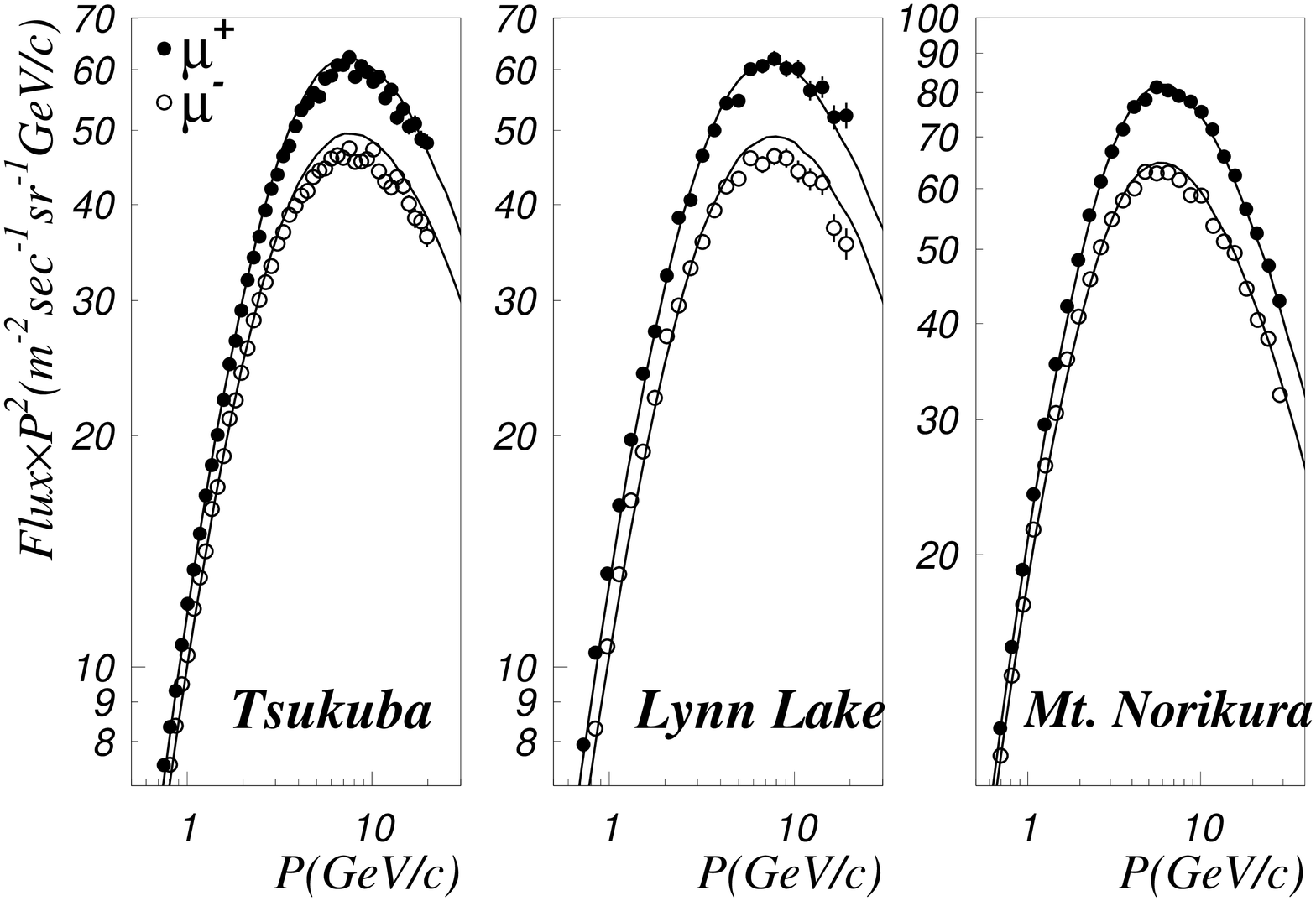}
  \end{center}
\vspace{-0.5cm}
  \caption{The observed and calculated muon spectra 
    at Tsukuba, Lynn Lake and  Mt. Norikura.}
  \label{fig:compar2}
\end{wrapfigure}

\section{Comparison of the data with calculation.}

We calculated the muon flux under the 
same environmental condition as that of the BESS-2001 balloon 
experiment, 
with several interaction models.
The primary flux model used here is essentially the one reported 
in Gaisser et al.\ [6] with a modulation function so 
that it reproduce the proton cosmic ray flux observed at 
4.68~${\rm g/cm^2}$  at the energies above the rigidity 
cut-off of 4.2~GV. 
The proton flux at 4.68~${\rm g/cm^2}$ calculated in this 
procedure is 
plotted in Fig.\ 2. 

We plot the quantity (muon flux)/depth for observed data 
in Fig.\ 4.
For the calculated flux, we depict the same quantity only for
4.68, 13.4 and 26.4~${g/cm^2}$ in the same figure for 
DPMJET-III~[11] and Fritiof~1.6~[8] (used in HKKM95) 
interaction models.
The curves for all atmospheric depths are very close.
It is clearly seen from these figures that the agreement between 
the data and calculation is better for DPMJET-III. Quantitatively,
the $\chi^2$ is 1.12 for DPMJET-III and 1.75 for Fritiof~1.6.
We have made the same analysis for FLUKA~97~[5] and 
Fritiof~7.02~[10],
and the $\chi^2$ are 1.37 and 1.85 respectively.
In Figs.\ 5, we plotted the ground level muon flux observed at Lynn lake,
Mt. Norikura, and Tsukuba. Also plotted are the muon flux calculated
by DPMJET-III. The muon fluxes at ground level are affected by 
the atmospheric density structure, and procedures
other than the hadronic interactions.
However, we can see that the
agreements of data and calculations are reasonably good.

In the comparison of data with calculated fluxes of the several 
interaction models, no interaction model is strongly excluded   
by the $\chi^2$ study. 
Among all the interaction models we studied here,
however, DPMJET-III is the most favored. \\

\section{Summary}
The BESS-2001 flight 
provided a very unique opportunity to measure precise cosmic-ray 
fluxes at small atmospheric depths
of 4.5 g/cm$^2$ through 28 g/cm$^2$. 
Using the primary and secondary cosmic-ray fluxes measured by the 
BESS-2001 experiment, we studied four interaction models used in the 
atmospheric neutrino calculations.
As a result of $\chi^2$ study, no interaction model was strongly 
excluded. However, DPMJET-III is the most favored 
in all the interaction models we studied here.
It reproduced atmospheric muons observed at sea level (Tsukuba and 
Lynn Lake) and mountain altitude (Mt. Norikura).

\vspace{\baselineskip}

We would like to thank ICRR, the University of Tokyo for the support.
This study was supported by Grants-in-Aid, KAKENHI(12047206), from
the Ministry of Education, Culture, Sport, Science and Technology 
(MEXT).

\vspace{\baselineskip}
\re
1.\ Abe K. \ et al. \ Phys.\ Lett.\ B \ 564 (2003) 8. 
\re
2.\ Abe K.\ et al.\ Proc.\ 28th ICRC (Tsukuba) HE2.4 (2003) 1463.
\re
3.\ Ajima Y. \ et al.\  Nucl. Instr. and Meth.\ A 443 (2000) 71.
\re
4.\ Asaoka Y. \ et al.\  Nucl. Instr. and Meth.\ A 416 (1998) 236.
\re
5.\ Battistoni, G.\ et al.\ astro-ph/0207035, unpublished.
\re
6.\ Honda M.\ et al.\ Proc.\ 28th ICRC (Tsukuba) HE2.4 (2003) 1415.
\re
7.\ Motoki M. \ et al.\ Astropart.\ Phys.\ 19 (2002) 113.
\re
8.\ Nilsson-Almqvist B.\ et al.\ Comp.\ Phys.\ Comm. 43 (1987) 387.
\re
9.\ Orito S.\ et al.\ 
in : Proc. ASTROMAG Workshop, KEK Report KEK87-19,  
eds. J. Nishimura, K. Nakamura, and A. Yamamoto  
(KEK, Ibaraki, 1987)p.111.
\re
10.\ Pi H.\ et al.\ Comp.\ Phys.\ Comm.\ 71 (1992) 173A
\re
11.\ Roeseler S.\ et al.\ SLAC-PUB-8740, hep-ph/0012252, unpublished.
\re
12.\ Sanuki T.\ et al.\ 2000 Astrophys.\ J.\ 545 (2000) 1135.
\re
13.\ Sanuki T.\ et al.\ Phys.\ Lett.\ B 541 (2002) 234.
\re
14.\ Shikaze Y. \ et al.\  Nucl.\ Instr.\ and Meth.\ A 455 (2000) 596.
\re
15.\ Yamamoto A. \ et al.\  Adv.\ Space Res.\ 14 (1994) 75.
\endofpaper
\end{document}